\documentclass[twocolumn]{aastex631}
\usepackage{amsmath}

\begin{document}

\title{Impacting Atmospheres: How Late-Stage Pollution Alters Exoplanet Composition}

\author[0009-0000-3475-7499]{Emilia Vlahos}
\affiliation{Department of Physics and Trottier Space Institute, McGill University, 3600 rue University, H3A 2T8 Montreal QC, Canada}

\author[0000-0003-1728-8269]{Yayaati Chachan}
\altaffiliation{CITA National Fellow}
\affiliation{Department of Physics and Trottier Space Institute, McGill University, 3600 rue University, H3A 2T8 Montreal QC, Canada}
\affiliation{Trottier Institute for Research on Exoplanets (iREx), Universit\'e de Montr\'eal, Canada}
\affiliation{Department of Astronomy and Astrophysics, University of California, Santa Cruz, CA 95064, USA}

\author[0000-0002-2870-6940]{Vincent Savignac}
\affiliation{Department of Physics and Trottier Space Institute, McGill University, 3600 rue University, H3A 2T8 Montreal QC, Canada}
\affiliation{Trottier Institute for Research on Exoplanets (iREx), Universit\'e de Montr\'eal, Canada}
\affiliation{Department of Astronomy \& Astrophysics, University of California, San Diego, La Jolla, CA 92093-0424, USA}

\author[0000-0002-1228-9820]{Eve J.~Lee}
\affiliation{Department of Physics and Trottier Space Institute, McGill University, 3600 rue University, H3A 2T8 Montreal QC, Canada}
\affiliation{Trottier Institute for Research on Exoplanets (iREx), Universit\'e de Montr\'eal, Canada}

\begin{abstract}

Atmospheric composition of exoplanets is often considered as a probe of the planet's formation condition. How exactly the initial chemical memory may be altered from the birth to the final state of the planet, however, remains unknown. Here, we develop a simple model of pollution of planetary atmosphere by the vaporization of infalling planetesimal of varying sizes and composition (SiO$_2$ inside 1 au and H$_2$O outside 1 au), following their trajectory and thermal evolution through the upper advective and radiative layers of a sub-Neptune class planet during the late stage of disk evolution. We vary the rate of pollution by changing the solid content of the disk and by dialing the level of disk gas depletion which in turn determines the rate of planetary migration. We find that pollution by silicate grains will always be limited by the saturation limit set by the thermal state of the atmosphere. By contrast, pollution by water ice can lead to $\sim$2--4 orders of magnitude variation in the atmospheric water mass fraction depending on the solid and gas content of the disk. Both cases suggest that post-formation pollution can erase the initial compositional memory of formation. Post-formation pollution can potentially transform sub-Neptunes with H/He-dominated envelope that initially formed beyond the iceline to waterworlds (water-enriched envelope) when the disk gas is depleted by $\gtrsim$2 orders of magnitude,  allowing gentle migration. We additionally discuss the expected C/O ratio profile under pollution by water and refractory carbon species.
\end{abstract}

\section{Introduction} \label{sec:intro}

One of the major goals of characterizing the chemical make-up of exoplanetary atmospheres is to infer the formation condition of the target planets. Relating a planet's composition to its formation, however, remains non-trivial as it depends on the physical and chemical evolution of protoplanetary disks in which planets are born \citep{Oberg2011, Oberg2021}, the planets' migration history, and details of their accretion process \citep[e.g.,][]{Molliere2022}. The complexity and interplay of these processes can often render inferences of formation history from composition non-unique. At its most simplest form, planets with atmospheres heavily metal-enriched with respect to the metallicity of the host star may indicate accretion of heavy element species in excess of the global mean value, which may arise from inheriting radially varying elemental ratios across evolving icelines \citep{Piso2015, Molliere2022} or late-stage pollution \citep{Kral2020}. Here, we focus our attention on pollution by accreting solids.

As the solids enter and fall through the planetary envelope, they are expected to ablate, like meteorites in the terrestrial atmosphere \citep{Love_1991, Zahnle1992} and the Shoemaker-Levy 9 impact on Jupiter \citep{Chevalier1994, MacLow1994}. When the planet is still embedded in a gas disk, the early envelopes are volumetrically large and hot allowing for solids to ablate before reaching the planetary cores. Both the observed metal enrichment of Jupiter in the solar system \citep{Pollack1996, DAngelo2021} and the creation of its diluted inner core \citep{Helled2017, Valletta2019} have been explained by the accretion and ablation of solids. Similar formation calculation has been computed for Neptune-like and sub-Neptune planets, focusing on the formation of central core \citep{Brouwers_2018} or the formation of a specific planetary system Kepler-36 \citep{Bodenheimer2018}.

More generally, on a population level, atmospheric measurements pre-dating James Webb Space Telescope hint at larger variation in the observed metallicity for lower-mass planets compared to higher-mass planets \citep[e.g.,][see their Figure 8]{Zhang2020}, which is expected from early population synthesis models \citep[e.g.,][]{Fortney2013}. Sub-Neptunes only possess a few \% by mass envelope \citep{Wolfgang_2015} and so it is natural to expect that their final atmospheric composition would be sensitively determined by the accretion history of solids and therefore display a much larger diversity in their metal enrichment compared to the gas giants (see also \citealt{Zlimen2024} for the case of Uranus and Neptune; \citealt{D_Angelo_2015} for a gas giant).

While prior studies of small planets have concentrated on early formation models, in this paper, we focus solely on the effect of pollution by solid accretion during the late stage of disk evolution after the planet has established its H/He-rich envelope. Our aim is to quantify how much post-formation processes can alter the initial atmospheric compositions set by the early formation of planets, over a wide range of disk dispersal history and the orbital location of the planet. 

The paper is organized as follows. We lay out our accretion and ablation model in Section~\ref{sec:model}. In Section~\ref{sec:results}, we show our results for the behavior of solid impactors in a planetary atmosphere and the extent of enrichment as a function of disk and planet properties that control the accretion rate. We discuss the implications and caveats of our work in Section~\ref{sec:discussion} and summarize our findings and conclude in Section~\ref{sec:conclusions}.

\section{Model}
\label{sec:model}

We begin with a full-fledged planet consisting of a 5$M_\oplus$ core and a gas-to-core mass ratio of 0.0591, representing a gas-enveloped mini-Neptune \citep[e.g.,][]{Wolfgang_2015}.\footnote{Such specific gas-to-core mass ratio is taken due to the design of our numerical calculation of planetary envelope structure. Our result is robust against the exact value of the envelope mass fraction, which we discuss in more detail in Section \ref{subsec:PlanetMass}.} As the planet orbits and/or migrates through an evolved disk of gas and solids, it is impacted by solid planetesimals over 2 Myr, the timescale over which the remaining disk gas dissipates away \citep{Mamajek_2009, Michel_2021}. The frictional and radiative heating of the impactors can evaporate them, polluting the local gas layer with the metallic content. Varying the composition of the impactor between pure silicate inside 1 au and pure water ice outside 1 au, we track the amount of evaporated material deposited into the atmospheric layers to determine how late-stage solid accretion alters atmospheric metallicity. 

In Section \ref{subsec:disk} we describe the properties of the background protoplanetary disk we adopt.
In Section \ref{ssec:accr-rate}, we outline the calculation of solid accretion rate. In Section \ref{subsec:atmosphere}
we describe the construction of planetary atmospheric profiles. In Section \ref{subsec:pollute} we outline the calculation of impactor evaporation within the envelope. Section \ref{subsec:Enrich} describes how metallicity is estimated from the planet's accretion history and the impactors' evaporation. How we put all the ingredients together is summarized in Section \ref{subsec:ModelSummary}.

\subsection{Disk Model}\label{subsec:disk}

Noting that the disk will be depleted by the initial formation of planets, we leave the disk density as a free parameter representing various stages of gas and solid depletion. We assume an initial disk surface density profile given by the minimum mass extrasolar nebula (MMEN) of \citet{Chiang_2013} and reduce the gaseous and solid disks independently by a factor of $f_{\text{gas}}$ and $f_{\text{solid}}$ respectively: 
\begin{equation}\label{eq:SolidSurfaceDensity}
    \Sigma_{\text{s}} =  6.2\times 10^2 f_{\text{solid}} \left(\frac{a}{0.2\text{au}}\right)^{-1.6}\,{\rm g\,cm^{-2}}
\end{equation}
and
\begin{equation}\label{eq:GasSurfaceDensity}
    \Sigma_{\text{g}} = 1.3\times 10^5 f_{\text{gas}} \left(\frac{a}{0.2\text{au}}\right)^{-1.6}\,{\rm g\,cm^{-2}}
\end{equation}
where $a$ is the orbital distance. We use $f_{\text{gas}} \in [10^{-2}, 10^{-4}]$ and $f_{\text{solid}} \in [10^{-4}, 10^{-5}, 10^{-6}]$, appropriate for the late stage of disk evolution where both the gas and solid are depleted.

We convert between surface density and volumetric density using the scale height of the disk, $H$:
\begin{equation}\label{eq:DiskDensity}
    \rho = \frac{\Sigma}{\sqrt{2 \pi}H},
\end{equation}
where in case of gas disk $H_g = c_s/\Omega$, $c_s = \sqrt{kT/\mu m_H}$ is the sound speed, $k$ is the Boltzmann constant, $\mu=2.37$ is the mean molecular weight, $m_H$ is the atomic mass of Hydrogen, $\Omega_{\rm k} = \sqrt{GM_\star/a^3}$ is the Keplerian orbital frequency, $G$ is the gravitational constant, and $M_\star$ is the stellar mass. In case of solid disk, we use \begin{equation}\label{eq:SolidScaleHeight}
    H_s = \frac{v_{\text{rel,s}}}{\Omega_{\text{k}}}.
\end{equation}
where $v_{\rm rel,s}$ is the relative velocity between solids.

We assume a disk temperature profile of
\begin{equation}\label{eq:DiskTemperature}
    T(a) = 1000\text{K} \times\left(\frac{a}{0.1 \text{au}}\right)^{-\frac{3}{7}},
\end{equation}
where the temperature normalization is taken from the the MMEN (\citealt{Chiang_2013}; see also \citealt{DAlessio98}) and the scaling is modified to account for stellar irradiation \citep[e.g.,][]{Chiang_1997}. 

\subsection{Mass Accretion Rate}\label{ssec:accr-rate}
\subsubsection{In Situ Accretion}\label{subsec:InSitu}
We consider first that the planet's orbital distance remains constant throughout accretion, resulting in a time-invariant solid accretion rate.
In general, the accretion rate is estimated by the flux of mass that impacts the planet's accretion cross section throughout its orbit. Our planets are still embedded in the gaseous nebula and so we take the accretion radius as the outer bound radius of the planetary envelope 
$R_{\rm acc} = R_{\rm out} = {\rm min}(R_{\rm Bondi}, R_{\rm Hill})$ where $R_{\rm Bondi} \equiv GM_p/c_s^2$ is the Bondi radius, $M_p$ is the mass of the planet, and $R_{\rm Hill}=(M_p/3M_\star)^{1/3}a$ is the Hill radius.
The accretion rate is given by
\begin{align}\label{eq:AccretionRate}
    \dot{M} &= \rho_{\text{s}}v_{\text{rel}}\pi R^2_{\text{acc}} \nonumber \\
    &= \sqrt{\frac{\pi}{2}}\Sigma_{\text{s}}\Omega_{\text{k}} R^2_{\text{acc}}
\end{align}
where $\rho_{\text{s}}$ is the volumetric density of solids (c.f.~Eqs. \ref{eq:DiskDensity} and \ref{eq:SolidScaleHeight}) in the disk and $v_{\text{rel}}$ is the relative velocity between the planet and solid impactors, which we assume is equal to the relative velocity between solids in the disk, $v_{\text{rel,s}}$, which is expected for planetesimals that are stirred by the planet \citep[e.g.,][]{Rafikov04}.

\subsubsection{Migration} \label{subsec:Migration}
As the planet sweeps through the disk, it interacts gravitationally with the surrounding gas, exchanging energy and momentum \citep{Goldreich_1980, Lin_1986}. Consequently, the planet may not remain in a fixed orbit but rather migrate radially. We investigate how migration affects solid accretion and subsequently atmospheric metallicity. 

In typical disks, the net torque on the planet by the perturbed gas can be approximated as \citep{Kley_2012}
\begin{equation}\label{eq:Torque}
    \tau = -(1.36 + 0.62\beta_\Sigma +0.43\beta_T)\frac{\text{G}^2M_p^2 \Sigma_{\text{g}}}{c_{\text{s}}^2},
\end{equation}
where $\beta_\Sigma = 1.6$ is the gas disk surface density power-law index, 
$\beta_T = 3/7$
is the temperature profile power-law index. The negative sign denotes that the force acts opposite to the planet's motion. The planet's migration rate is determined by equating Eq.~\ref{eq:Torque} to the time derivative of the planet's angular momentum, $L = M_p\Omega_{\text{k}}a^2$, and solving for its radial velocity:
\begin{equation}\label{eq:MigrationRate}
    \dot{a} = \frac{2\tau}{M_p\Omega_{\text{k}} a}.
\end{equation}

We solve for $a(t)$ analytically by integrating Eq.~\ref{eq:MigrationRate} over time until $a(t)$ reaches the innermost edge of the disk, 0.1 au. When calculating accretion, we divide the accretion timescale into discrete timesteps of 10 years, $t_i$, with associated positions $a_i = a(t_i)$. Within a given timestep, the accreted mass is determined by evaluating Eq.~\ref{eq:AccretionRate} at $a_i$. For simplicity, we keep the planet's atmosphere (discussed in the next Section) static throughout migration. We consider planets migrating inwards from 3.5 au and vary $f_{\text{gas}} \in [1, 10^{-2}, 10^{-4}]$ allowing for rapid, moderate, and slow migration, respectively, for our chosen disk gas lifetime taken as 2 Myr.

\subsection{Atmospheric Model}\label{subsec:atmosphere}
The evaporation of accreted materials depends on the temperature, pressure, and density profile of the planetary envelope. 
We generate atmospheric profiles following the one-dimensional thermodynamic model of \cite{Lee_2014} updated to account for outer advective flows as derived in \citet{Savignac_2023}. Here, we only summarize the essential ingredients. We construct a spherically symmetrical hydrostatic envelope of gas-to-core mass ratio 0.0591 atop 5$M_\oplus$ core embedded in the gaseous nebula of Eq.~\ref{eq:GasSurfaceDensity} at 0.1, 1, and 3.5 au for dusty and dust-free opacity for a fiducial initial atmospheric metallicity of $Z=0.02$ (see Section \ref{subsec:HighMetallicities} for varying initial metallicity), hydrogen mass fraction 0.7 and helium mass fraction 0.28. The metallic species elemental abundance follows that of \citet{GN93}. The depth of the advective flow is set to 0.3 following typically quoted values in hydrodynamic simulations \citep[e.g.,][]{Lambrechts17,Bailey_2023}.

Our atmospheres follow a general three-layer structure of convective, radiative, and advective layer from inside-out. The outermost advective layer is not considered ``bound'' to the planet and within the innermost convective layer, vigorous mixing is expected to homogenize the metallicity and so we are only interested in the pollution in the sandwiched radiative layer. Nevertheless, the outer advective layer sets the thermal boundary condition of the radiative zone, and the advective layer is computed with a modified mixing length theory which determines whether a cycling gas parcel is able to mix (i.e. cool down) within the envelope before being ejected back out into the disk \citep[see][their Section 2]{Savignac_2023}. 

\subsection{Pollution Model} \label{subsec:pollute}
For a given atmospheric profile, we calculate the trajectory and ablation of impacting solids, assuming head-on impact. The impactors are modeled as uniform spheres of either 10 m or 1000 m in initial radius, representative of planetesimals.\footnote{We do not consider pebble accretion because pebbles evaporate entirely in the outermost advective region of the envelope that would cycle out of planet and do not pollute the bound radiative zone.} 
The composition of the impactors is varied from pure silicate SiO$_2$ inside 1 au and pure water ice H$_2$O beyond.
Starting from the entry velocity of $v_{\rm rel}$ at $R_{\rm acc}$ from the core, we solve for the impactor's equation of motion
\begin{equation}
    \ddot{r} = -\frac{GM_{\rm core}}{r^2} + \frac{F_{\rm D}}{M_s}
    \label{eq:ImpactorEOM}
\end{equation}
where $r$ is the location of the impactor from the center of the planet, $M_{\rm core}$ is the core mass of the planet, $M_s$ is the mass of a single impactor, and $F_{\rm D}$ is the drag force which we describe in detail in the next section. The negative sign points towards the core.

The initial infall speed is chosen as the escape speed from the planet's potential at $R_{\rm out}$, which effectively assumes maximum speed for an impactor that is guaranteed to be accreted onto the planet. With our choice, we are tracking the ablation of the impactor through the outer advective zone within which gas is expected to escape the planet's potential within an orbital time. Our chosen impactor sizes are large enough to be decoupled from the gas advective flow. For instance, with the initial infall velocity of escape velocity from $R_{\rm out}$ of the planet, the characteristic infall time $R_{\rm out}/(2GM_p/R_{\rm out})^{1/2} \sim 0.4 \Omega^{-1}$ so that we may reasonably consider the impactors to fall through the planet's potential well with minor deflection by the gas advection.

\subsubsection{Drag Force} \label{subsubsec:drag}
 The drag force is given by 
\begin{equation}\label{eq:dragforce}
F_{\text{D}} = \frac{1}{2} C_{\text{D}}A_{\text{s}}\rho_{\text{g}} v_{\text{s}}^2,
\end{equation}
where $A_{\rm s}$ is the cross sectional area of the impactor, $\rho_{\text{g}}$ is the density of the gas, $v_{\text{s}}$ is the speed of the impactor, and $C_{\text{D}}$ is the drag coefficient. The drag coefficient encodes most of the physics and depends on the Reynolds number, $\text{Re} = R_{\text{s}}\rho_{\text{g}}\frac{v_{\text{s}}}{\eta_{\text{g}}}$, and the Mach number, $\text{Ma} =\frac{v_{\text{s}}}{c_{\text{s}}}$, where $R_{\text{s}}$ is the impactor radius, and $\eta_{\text{g}}$ is the dynamical viscosity of the gas. These dimensionless parameters characterize the drag regime of the interaction. For our model to be applicable over a range of formation conditions, we must express the drag in general terms. We thus adopt the prescription of \citet{Melosh_2008}, who performed an extensive review of the literature to derive a continuous expression of $C_{\text{D}}$ accurate across a wide array of drag regimes:
\begin{equation} \label{eq:dragcoeff}
C_D = 2 + (C_1-2)e^{-3.07\sqrt{\gamma_{\text{g}}} \mathcal{K} G_D} + C_2e^{\frac{-1}{2\mathcal{K}} },
\end{equation}
where $\mathcal{K} \equiv \frac{\text{Ma}}{\text{Re}}$, 
\begin{equation} \label{eq:StokesCoeff}
C_1 = \frac{24}{\text{Re}}(1 + 0.15\text{Re}^{0.678}) + \frac{0.407\text{Re}}{\text{Re} + 8710},
\end{equation}
\begin{equation}
\text{log}_{10}G_D = \frac{2.5(\text{Re}/312)^{0.6688}}{1 + (\text{Re}/312)^{0.6688}},
\end{equation}
\begin{equation}
C_2 = \frac{1}{\sqrt{\gamma}\text{Ma}}\left(\frac{4.6}{1 + \text{Ma}} + 1.7\sqrt{\frac{T_{\text{s}}}{T_{\text{g}}}}\right),
\end{equation}
$T_{\text{s}}$ is the surface temperature of the impactor, $T_{\text{g}}$ is the temperature of the surrounding gas, and $\gamma = 1/ (1 - \nabla_{\text{ad}})$ is the adiabatic index where the adiabatic gradient,
\begin{equation}
 \nabla_{\text{ad}} = -\left.\frac{\partial \text{log} S_{\text{g}}}{\partial \text{log} P_{\text{g}}}\right|_{T_{\text{g}}} \left(\left.\frac{\partial \text{log} S_{\text{g}}}{\partial \text{log} T_{\text{g}}}\right|_{P_{\text{g}}}\right)^{-1}
\end{equation}
is evaluated at the local gas temperature, pressure ($P_{\text{g}}$), and entropy ($S_{\text{g}}$) from the atmospheric models described in Section \ref{subsec:atmosphere}.

The above expression reduces to that for Epstein drag in the limit $\mathcal{K}\gg1$, and that of Stokes drag in the limit $\mathcal{K}\ll1$. The impacting solids that we model are sufficiently large for the interaction to be well within the Stokes drag limit. In this case, the drag coefficient above reduces to $C_{\text{D}} \approx C_1$ and is characterized by the Reynolds number. $C_1$ (Eq.~\ref{eq:StokesCoeff}) is the expression proposed by \citet{D_Angelo_2015}, who extend the original expression of \citet{Melosh_2008} to fit results from an empirical review by \citet{Brown_2003}. For small, slow impactors with  $\text{Re} \lesssim 1$, we recover Stokes's law, $C_{\text{D}} \approx 24/\text{Re}$. For $\text{Re} \gg 1$, we recover $C_{\text{D}} \approx 4.07$, the Newtonian drag coefficient \citep{Whipple_1973}. Eq.~\ref{eq:StokesCoeff} is found to fit empirical data for Re below a critical Reynolds number of approximately $3\times 10^5$ \citep{Brown_2003}. For supercritical Renoylds numbers, we follow the suggestion of \citet{Brouwers_2018}, and model the drag coefficient of the turbulent flow by a step function decrease to $C_{\text{D}} = 0.2$ \citep{Michaelides_2006, Perry_1997}.

\subsubsection{Surface Temperature and Ablation} \label{subsec:ablation}
Within the envelope, gas drag gives rise to frictional heating of the impactor's surface. The increasingly hot gas surrounding the impactor as the latter falls towards the deeper atmosphere further contributes to its heating. As the impactor's surface temperature increases, it evaporates, enriching the surrounding atmosphere. We model the thermal ablation of the impactor according to the Knudsen–Langmuir equation which gives the mass vaporization rate as
\begin{equation} \label{eq:langmuir}
\dot{M_{\text{s}}} = -4\pi R_{\text{s}}^2P_{\text{v}}\sqrt{\frac{\mu_{\text{s}}m_H}{2\pi kT_{\text{s}}}}
\end{equation}
where $M_{\text{s}}$ is the impactor mass, $\mu_{\text{s}}$ is the mean molecular weight of the impactor species (59.6 for SiO$_2$ and 17.9 for H$_2$O), and $P_{\text{v}}$ is the temperature dependant vapour pressure of the impactor. 

In general, vapor pressure is determined by fitting empirical data to the integrated Clausius–Clapeyron equation (the Antoine Equation), and including necessary extensions. For SiO$_2$, we use
\begin{equation} \label{eq:VaporPressureSiO2}
\text{ln} P_{\text{v}} = a_0 - \frac{a_1}{T_{\text{s}}+a_2}
\end{equation}
where $a_0, \, a_1, \, a_2$ are best fit parameters provided by the NIST Chemistry WebBook based on experimental data from \citet{Stull_1947}. Note that the data is fit over a limited range of temperatures and should be considered an extrapolation for temperatures under $\sim$2000K, as in \citet{D_Angelo_2015} and \cite{Brouwers_2018}. 

For H$_2$O below its melting point ($T_{\text{s}} < 273.15$ K), we use the vapor pressure equation of \cite{Washburn_1924} wherein
\begin{equation} \label{eq:VaporPressureH2O_ice}
\text{log}_{10}P_{\text{v}} = a_0T_{\text{s}}^{-1} + a_1 + a_2T_{\text{s}} + a_3T_{\text{s}}^{2} + a_4\text{log}_{10}T_{\text{s}}.
\end{equation}
While the corresponding fit parameters shown in Table \ref{tab:VaporPressureConstants} are derived from vapor pressure data ranging from $\sim$173--273K, \citet{D_Angelo_2015} find the same parameters to also agree with the more recent data of \citet{Haynes_2011} extending down to 50K. 
For temperatures above the melting point and below the critical temperature of water, $T_{\text{cr}} =  647.096$K, we take the formula of \citet{Pruß_1995}: 
\begin{eqnarray}
\label{eq:VaporPressureH2O_water}
\text{ln}\left(\frac{P_{\text{v}}}{P_{\text{cr}}}\right) = \left(\frac{T_{\text{s}}}{T_{\text{cr}}}\right)(a_0\theta + a_1\theta^{1.5} + a_2\theta^3 + a_3\theta^{3.5} \nonumber\\
+ a_4\theta^4 + a_5\theta^{7.5}),
\end{eqnarray}
where $\theta = (1 - T_{\text{s}}/T_{\text{cr}})$ and $P_{\text{cr}} = {2.2064\times 10^8}$ dyne/cm$^2$ is the vapor pressure at the critical temperature. All coefficients $a_{\text{i}}$ for Eqs.~\ref{eq:VaporPressureSiO2} - \ref{eq:VaporPressureH2O_water} are given in Table~\ref{tab:VaporPressureConstants} with vapor pressure expressed in dyne/cm$^2$.

\begin{table}[]
\begin{flushleft}
\begin{center}
\begin{tabular}[t]{lllllll}
\hline
Eq. & $a_0$ & $a_1$ & $a_2$ & $a_3$ & $a_4$ & $a_5$\\
\hline
\ref{eq:VaporPressureSiO2} & 31.8 & 46071.4 & 59.8 & --- & --- 
& ---\\
\ref{eq:VaporPressureH2O_ice} & -2445.6 & -3.63 & -0.0168 & 0.0000121 & 8.23 & ---\\
\ref{eq:VaporPressureH2O_water} & -7.86 & 1.84 & -11.8 & 22.7 & -16.0 & 1.80\\
\hline
\end{tabular}
\caption{Fit parameters for vapor pressure formulas of SiO$_2$ and H$_2$O above and below melting point. All vapor pressures have units dyne/cm$^2$.}
\label{tab:VaporPressureConstants}
\end{center}
\end{flushleft}
\end{table}

To obtain the vaporization rate of the impactor, we solve for its surface temperature evolution. The primary heating mechanisms acting on the impactor are frictional dissapation and thermal radiation. As the impactor slows due to drag within the envelope, a fraction ($f$) of its kinetic energy is dissipated in the form of heat. The associated power is given by
\begin{equation}\label{eq:FrictionalPower}
P_{\text{fric}} = fF_{\text{D}}v_{\text{s}}.
\end{equation}
To estimate the factor $f$, we follow the approach of \citet{Melosh_2008} which allows $f$ to vary between different drag regimes:
\begin{equation}
f = \frac{8}{\gamma_{\text{g}}} \left( \frac{\text{Nu}}{\text{RePr}} \right) \frac{\text{Pr}^{\frac{1}{3}}}{C_{\text{D}}},
\end{equation}
where Nu is the Nusselt number
\begin{equation}
\text{Nu} = \frac{\text{Nu}_{\text{C}}}{1 + (3.42\text{M}'\text{Nu}_{\text{C}}/\text{RePr})},
\end{equation}
\begin{equation}
\text{Nu}_{\text{C}} = 2 + 0.459\text{Re}^{0.55}\text{Pr}^{\frac{1}{3}},
\end{equation}
\begin{equation}
\text{M}' = \frac{\text{Ma}}{1+0.428\text{Ma}(\gamma+1)/\gamma},
\end{equation}
and Pr is the Prandtl number
\begin{equation}
\text{Pr} = \frac{\eta_{\text{g}}C_{\text{p,g}}}{k_{\text{g}}},
\end{equation}
where $C_{\text{p,g}} = \frac{P_{\text{g}}}{\rho_{\text{g}} T_{\text{g}} \nabla_{\text{ad}}}$ is the specific heat of the gas, and $k_{\text{g}}$ the thermal conductivity of the gas, for which we use the temperature dependant thermal conductivity of hydrogen, 
\begin{equation}
k_{\text{g}} = 41.84 \,{\rm erg\,s^{-1}\,cm\,K} \left[1434 + 1.257 \cdot \left(\frac{T}{\rm K} - 1200\right)\right],
\end{equation}
based on the fit provided by \citet{Normand_1960}.

Thermal radiation is driven by the temperature difference between the impactor and the atmospheric gas. Generally, the gas will be hotter than the impactor's surface and will irradiate energy to the impactor. However, if the impactor undergoes sufficient friction and insufficient ablation, its surface can transfer excess heat into the atmosphere, which we observe in our models for pollution by quartz in the outer envelope. The radiative power into or out of the impactor is given by,
\begin{equation}\label{eq:RadiativePower}
P_{\text{rad}} = 4\pi r_{\text{s}}^2\alpha_{\text{s}} \sigma_{\text{sb}} (T_{\text{g}}^4-T_{\text{s}}^4),
\end{equation}
where $\alpha_{\text{s}}$ is the absorbtivity of the impactor, assumed to be 1, and $\sigma_{\text{sb}}$ is the Stephan-Boltzmann constant.

As the impactor's surface heats, it will shed mass following Eq.~\ref{eq:langmuir}. The process of vaporization requires energy and will therefore cool the impactor. The power associated with this evaporative cooling is given in \citet{Brouwers_2018} as
\begin{equation}\label{eq:EvaporativePower}
P_{\text{evap}} = - E_{\text{vap}} \cdot \dot{M_{\text{s}}},
\end{equation}
where $E_{\text{vap}}$ is the latent heat of vaporization of the solid, taken as $8.08 \times 10^{10}$ erg/g for quartz and $2.83 \times 10^{10}$ erg/g for water ice as in \citet{D_Angelo_2015}. 

The sum of the power terms in Eq~\ref{eq:FrictionalPower}, Eq~\ref{eq:RadiativePower}, and Eq~\ref{eq:EvaporativePower} gives the change in the internal (thermal) energy of the impactor. We find this sum to be negligible (see also \citealt{Podolak_1988}) and so we set
\begin{equation}\label{eq:SurfaceTemperature}
P_{\text{rad}} + P_{\text{fric}} + P_{\text{evap}} = 0,
\end{equation} 
to evaluate $T_s$ at each timestep. Physically, the energy gained through friction and radiation serves primarily to evaporate the impactor and does not contribute significantly to its net thermal energy.

\subsubsection{Shock}
It is common for impactors to reach supersonic velocities during their infall. The shock-heated gas layer is in direct contact with the impactor and is expected to contribute to the latter's vaporization. The thermal state of the shocked layer depends on whether the shock is adiabatic (inefficient cooling) or isothermal (efficient cooling). We determine this thermal state by comparing the timescale over which the impactor falls through the shock wave, $t_{\text{infall}}$, to the timescale over which heat diffuses out of the shock wave, $t_{\text{diff}}$. We evaluate the infall timescale according to 
\begin{equation} 
t_{\text{infall}} = \frac{d_{\text{sh}}}{v_{\text{s}}},
\end{equation}
where $d_{\text{sh}} = \lambda_{\text{mfp}} / \text{Ma}$ is the width of the shock wave and $v_{\text{s}}$ is the impactor velocity. We calculate the mean free path, $\lambda_{\text{mfp}} = \mu_{\text{g}}m_H/ \pi d_{\text{g}}^2\rho_{\text{g}}$, approximating $\mu_{\text{g}} = 2$ and $d_{\text{g}} = 2.9\times 10 ^{-8}$ cm as the mean molecular weight and diameter of diatomic hydrogen molecule. We evaluate the diffusion timescale according to 
\begin{equation} 
t_{\text{diff}} = \frac{C_{\text{p,g}} M_{\text{{sh}}} \Delta T}{A_{\text{s}}\sigma_{\text{sb}} T_{\text{sh}}},
\end{equation}
where $M_{\text{{sh}}}$ is the gas mass in the shocked layer, $\Delta T$ is the temperature difference pre/post shock, $T_{\text{sh}}$ is the temperature of the shocked gas, and we model the cross-section of the shocked layer to be that of the impactor. We find that the ratio $t_{\text{infall}}/t_{\text{diff}}$ is generally orders of magnitude less than unity, indicating that the shock is predominantly adiabatic. We therefore calculate the post-shock temperature, pressure, and density using the jump conditions for adiabatic shock:
\begin{equation} 
T_{\text{sh}} = T_{\text{g}}\cdot\frac{\left(2\gamma \text{Ma}^2 - (\gamma-1)\right)\left((\gamma-1)\text{Ma}^2 + 2\right)}{(\gamma+1)^2\text{Ma}^2},
\end{equation}
\begin{equation} 
P_{\text{sh}} = P_{\text{g}}\cdot\frac{\left(2\gamma \text{Ma}^2 - (\gamma-1)\right)}{\gamma+1},
\end{equation}
\begin{equation} 
\rho_{\text{sh}} = \rho_{\text{g}}\cdot\frac{(\gamma+1)\text{Ma}^2}{(\gamma-1)\text{Ma}^2 + 2},
\end{equation}
where $\gamma$ is the local gas adiabatic index, and we use $T_{\rm sh}$, $P_{\rm sh}$, and $\rho_{\rm sh}$ in Eq.~\ref{eq:FrictionalPower} and \ref{eq:RadiativePower}.

\subsection{Enrichment and Rain Out}\label{subsec:Enrich}
Section~\ref{subsec:pollute} determines the amount of material deposited into the envelope by a single impactor. The net pollution is the sum of the material deposited by all impactors accreted over a 2 Myr, where the total impactor accretion rate is deduced from the mass accretion rate (Section \ref{ssec:accr-rate}).
We calculate the metallicity (evaporated mass / total gas mass) as a function of radius within the envelope after an accretion timescale of 2 Myr and compare this to the initial baseline atmospheric metallicity ($Z=0.02$ in the fiducial; see Section \ref{subsec:HighMetallicities} for other cases). 

The upper limit on the amount of mass that can be deposited into the atmosphere before saturation (relative to the initial gas mass) is set by the vapour pressure curve of the accreted material:
\begin{equation}\label{satsat}
f_{\text{sat}} = \frac{\mu_{\rm s}}{\mu_{\text{g}}}\frac{P_{\rm v}(T_{\text{g})}}{P_{\text{g}}},
\end{equation}
where $\mu_{\text{g}}$ is the mean molecular weight of the atmospheric gas at a given altitude.
Beyond saturation, we assume that any additional deposited mass will rain out deep into atmospheric layers beyond the radiative region of the atmosphere; we do not consider how rained out material might revaporize. Note that in the case of water, $P_{\rm v}(T_{\text{g}})$ can only be evaluated for gas temperatures less than the critical temperature of 647.096K. At higher temperatures, H$_2$O can exist in the envelope in any mixing ratio, and we therefore leave it unbound.

\begin{table}[]
\begin{flushleft}
\begin{center}
\begin{tabular}[t]{lll}
\hline
\textbf{Description} & \textbf{Parameter} & \textbf{Values}\\
\hline
Core Mass ($\text{M}_{\oplus}$) & $M$ & 5\\
Advection depth  & $\alpha_{\text{ad}}$ & 0.3\\
Opacity & & Dusty\\
Orbital Distance (au) & $a$ & 0.1, 3.5\\
Solid Disk Depletion & $f_{\text{solid}}$ & $10^{-4}$, $10^{-5}$, $10^{-6}$\\
Gas Disk Depletion & $f_{\text{gas}}$ & (1),  $10^{-2}$, $10^{-4}$\\
Impactor Radius (m) & $r_{\text{s}}$ & 10, 1000 \\
Solid Disk Composition & & SiO$_2$ for $a < 1$au \\
& &  H$_2$O for $a > 1$au\\
\hline
\end{tabular}
\caption{Description and values of input parameters across accretion models. $f_{\text{gas}}$ = 1 is only considered for migrating planets, to include rapid migration rates. For in situ accretion, we vary orbital distance, solid/gas disk depletion, and impactor size for a total of 24 models. For migration, all models begin at an initial orbital distance of 3.5 au, and we vary the remaining parameters for a total of 18 models. We omit the dust-free profiles because the results overlap largely with the dusty profiles, with the exception of the $f_{\text{gas}} = 10 ^{-4}$ envelope at 3.5 au for which the dust-free profile is notably colder than the dusty profile, lowering the saturation curve of SiO$_2$ by approximately 4 orders of magnitude.}
\label{tab:InputParameters}
\end{center}
\end{flushleft}
\end{table}

\begin{figure*}[]
  \centering
\includegraphics[width=1\textwidth]{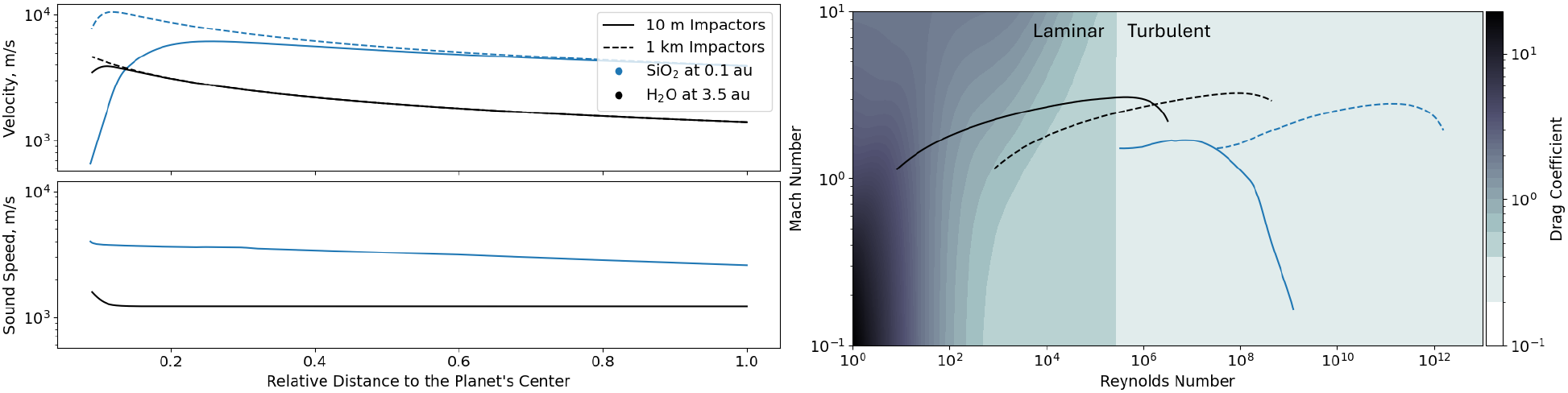}
\caption{The left panels show the infall velocity (top) and atmospheric sound speed (bottom) for impactors of 10m (solid) and 1000m (dashed) and at 0.1 au (blue) and 3.5 au (black). The x-axis represents the distance to the planet's center, normalized with respect to the outer radius ($r / R_{\text{out}}$). The right panel shows the drag regimes of these 4 cases as characterized by their Mach/Reynolds numbers. The corresponding drag coefficient (Eq.\ref{eq:dragcoeff}) is shown in the contour plot. In the Epstein regime (Ma$\gg$Re; bottom-left corner), the drag coefficient depends on the ratio between the impactor surface temperature and the gas temperature. Here, we consider the case $T_{\text{s}}=T_{\text{g}}$. All atmospheric profiles have dusty opacity and gas disk densities of 0.01 MMEN.}
\label{fig:DragParameters}
\end{figure*}

\subsection{Model Summary}\label{subsec:ModelSummary}
Table \ref{tab:InputParameters} summarizes the parameters we use. We solve numerically for the evolution of a single impactor into a planet envelope using a Runge-Kutta-Fehlberg scheme with an adaptive stepsize to update the impactor's position, velocity, and mass. 
We evaluate its trajectory by considering the opposing forces of gravity and gas drag (Eqs.~\ref{eq:ImpactorEOM} and \ref{eq:dragforce}) and evaluate its evaporation according to Eq.~\ref{eq:langmuir}. 
At each timestep the impactor's surface temperature is updated according to Eq.~\ref{eq:SurfaceTemperature}. We calculate the rate at which impactors of a given size are accreted from Eq.~\ref{eq:AccretionRate}. For each impactor accreted, we determine the amount of mass deposited into the atmosphere by the impactor's vaporization curve and update metallicity accordingly. All models are evolved over 2 Myr. We report the deposition profile only down to the radiative-convective boundary because the memory of the pollution is likely erased by the vigorous mixing in the inner convective zone. In cases where there are multiple convective zones, we stop the integration at the uppermost radiative-convective boundary. In other words, all the calculations discussed and shown in this paper (including all our figures) are only down to the uppermost radiative-convective boundary.

\section{Results}
\label{sec:results}

\begin{figure*}[]
  \centering
\includegraphics[width=1\textwidth]{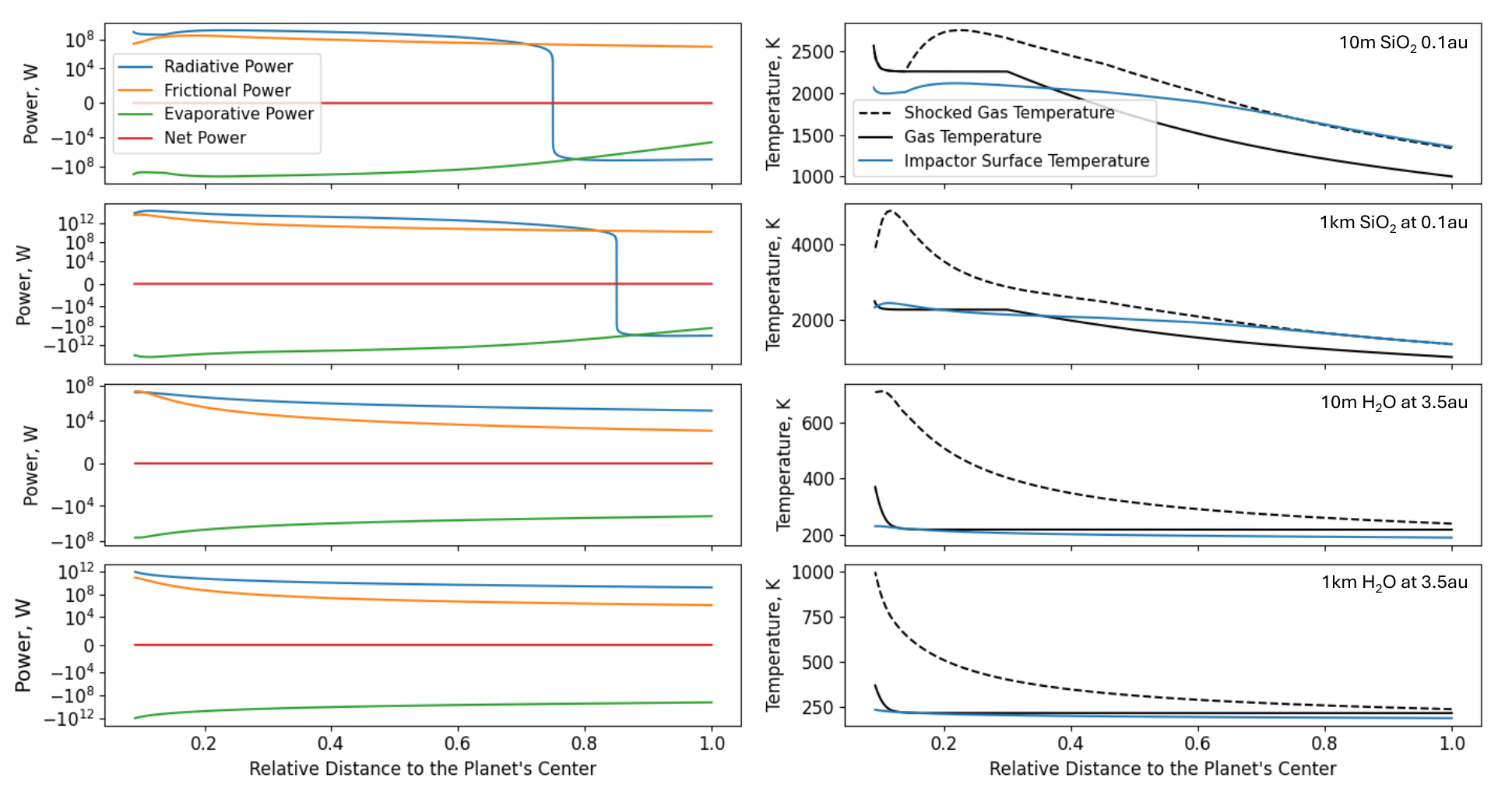}
\caption{Left: radial profiles of power terms that contribute to the thermal evolution of the impactor, as well as the net power (always zero according to Eq.~\ref{eq:SurfaceTemperature}). Right: the surface temperature of the impactor as well as the pre and post shock temperatures of the gas over the impactor trajectory. Impactor size and orbital distance is consistent between left and right panels for a given row. All atmospheric profiles have dusty opacity and $f_{\rm gas}=10^{-2}$.}
\label{fig:Thermal Evolution}
\end{figure*}

\begin{figure*}[]
  \centering
\includegraphics[width=1\textwidth]{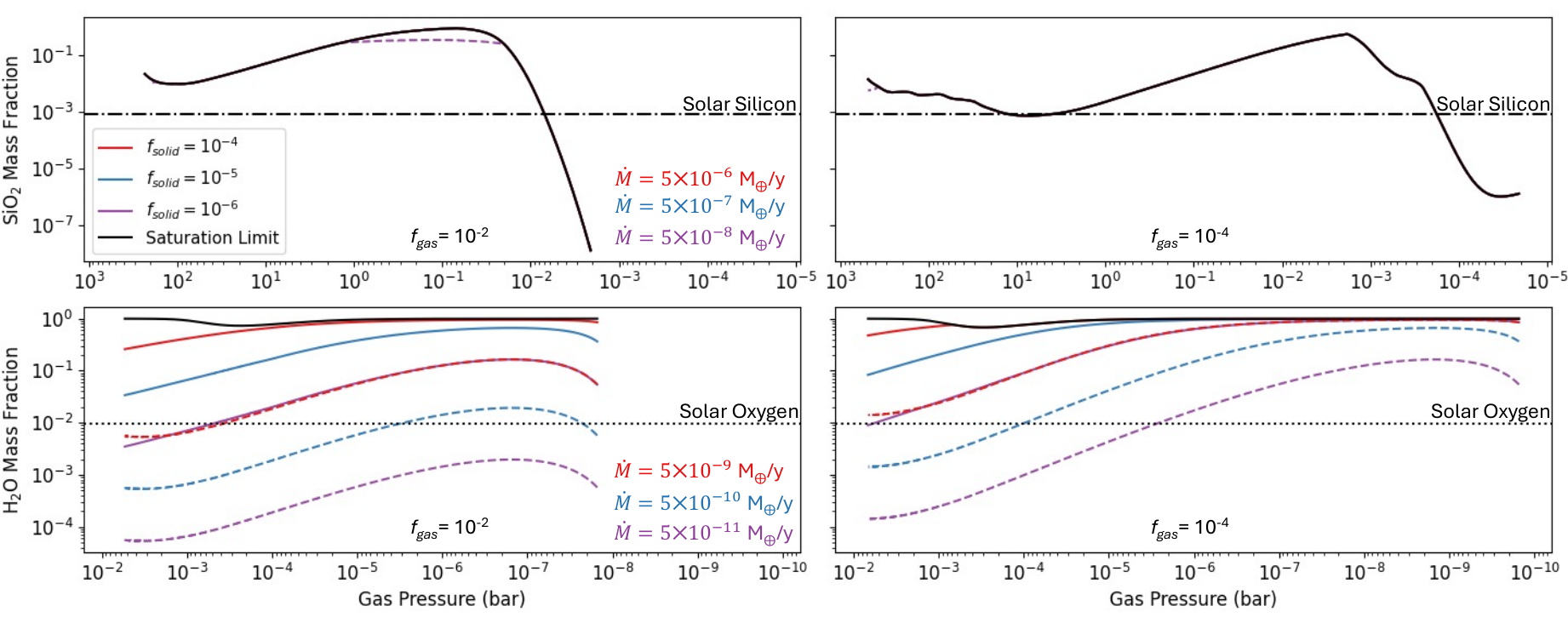}
\caption{Mass fractions of the accreted SiO$_2$ (at 0.1 au, top) and H$_2$O (at 3.5 au, bottom) relative to the to the total gas (including the evaporated mass) after an accretion timescale of 2 Myr. Solid lines with color represent the accretion of 10 m impactors and dashed lines with color represent the accretion of 1 km impactors. Different colors represent variations in solid disk density and the associated mass accretion rate at a given orbital distance. SiO$_2$ mass fraction is limited by saturation (black solid line), causing overlap between curves. By contrast, H$_2$O mass fraction is limited to unity in the case where the mass of deposited H$_2$O exceeds that of the initial gas. The left panels depict envelopes in gaseous disk densities of $10^{-2}$ MMEN and the right panels depict envelopes in gaseous disk densities of $10^{-4}$ MMEN. All profiles are for dusty opacities. We include solar silicon and oxygen abundances consistent with $Z=0.02$ and \citet{GN93} for comparison.}
\label{fig:In Situ Profiles}
\end{figure*}

\subsection{Drag Regimes}

The velocity of the impactor depends primarily on the envelope profile through which it falls. 
The entry velocity of impactors is higher at closer-in distances because smaller $R_{\rm out}$ boosts the escape velocity there.
Upon entering the atmosphere, impactors accelerate as they approach terminal velocity. The amount of acceleration is comparable between close-in and far-out planets since the increased gravitational force in close-in planets is counteracted by the increased drag force due to higher gas density. Velocities are therefore typically higher throughout the bulk of the atmospheres of close-in compared to far-out planets, until the impactors reach local terminal velocity and begin to decelerate as they descend into denser region, as seen in the upper left panel of Fig.~\ref{fig:DragParameters}. 
The point of deceleration occurs higher up in the atmosphere at close-in distances because the underlying gas density is generally higher there (see blue vs.~black curves in Fig.~\ref{fig:DragParameters}). At a given orbital distance, larger impactor requires higher gas density to reach its terminal velocity so we see their point of deceleration appears deeper in the atmosphere.

The right panel of Fig.~\ref{fig:DragParameters} shows the drag regime of impactors at varying orbital distance as a function of their Mach and Reynolds number. Envelopes of hot, close-in planets have higher sound speeds than those of cold, far-out planets (Fig.~\ref{fig:DragParameters} bottom left). This offsets the high impactor velocities, and we find Mach numbers of the outer envelope to be comparable across orbital distances, typically ranging between 1 and 3. The Mach number only decreases below 1 past the impactor's point of deceleration. The flow is therefore supersonic within the majority of the envelope in volume, aiding the vaporization of impactors even more. 

Due to high velocities and gas density, the Reynolds number is greatest for flow through close-in planets. At 0.1 au, we find a supercritical Reynolds number for both 10 m and 1 km impactors, indicative of turbulent drag. The drag coefficient in this domain is fixed at 0.2. At 3.5 au, Reynolds numbers are lower; drag is initially laminar (in the outer envelope) and becomes turbulent as the impactors accelerate. In both cases, Re $\gg$ 1 and Ma $\ll$ Re (equivalently $R_s \gg \lambda_{\rm mfp}$) so impactors are in the pressure drag regime. 

\subsection{Impactor Thermal Evolution}
Fig.~\ref{fig:Thermal Evolution} shows the thermal evolution of 10 m and 1 km impactors at 0.1 au (quartz) and 3.5 au (water ice). At 0.1 au for quartz impactors, the frictional power dominates initially because the relatively low vapor pressure of quartz prevents them from ablating significantly in the outer envelope. With minimal evaporation, quartz cannot cool via mass shedding and the surface temperature increases beyond that of the surrounding gas. At $\sim$0.8$R_{\rm out}$, the quartz impactor is now hot enough to evaporate; the evaporative cooling overtakes frictional heating and we see the impactor surface temperature falling below the shocked gas temperature and radiative power turning positive (i.e., heating).

Icy impactors ablate at much lower temperatures and generally undergo significant evaporation upon entering the envelope. They stay relatively cool compared to the local gas and gain energy from both frictional and radiative heating. Radiation is typically the dominant form of heating; however, deep in the atmosphere, gas density increases at a faster rate than gas temperature, and frictional heating becomes comparable to radiative heating. We see larger evaporative power for 1 km impactor because the amount of evaporation at a given temperature depends on impactor size, scaling with surface area as dictated by Eq.~\ref{eq:langmuir}. 

Recall that we do not account for the internal heating of the impactor. When frictional heating exceeds evaporation, we assume that thermal energy radiates back into the gas, rather than heating the impactor. When evaporation exceeds frictional heating, we assume that the outer layer of the impactor evaporates before heat can be conducted through it. In both cases, we expect the conduction timescale to be long relative to the cooling timescale (radiation or evaporation), which prevents the impactors interior from heating significantly. Furthermore, we performed a first order estimate of the power lost due to internal conduction \citep[see][]{D_Angelo_2015, Brouwers_2018} and found it to have negligible effect on the impactor's surface temperature. 

\subsection{In Situ Accretion}

The metallicity profiles for planets accreting in-situ at 0.1 au and 3.5 au are shown in Fig.~\ref{fig:In Situ Profiles}, top and bottom rows, respectively. We remind the readers that throughout our calculations, our planets remain embedded in the gaseous nebula so that our SiO$_2$ accreting envelopes, placed inside 1 au, will always have higher envelope gas pressure compared to our H$_2$O accreting envelopes, which are placed outside 1 au where the nebular temperature and density are both lower and so the nebular pressure is lower.
Atmospheres at 0.1 au typically become saturated in SiO$_2$ within the 2 Myr of accretion timescale (metallicity curves overlap with saturation limit). We conclude that the vapor pressure curve sets the silicate content of the radiative layers for planets forming close-in. In this sense, SiO$_2$ profiles are dependent on the envelope profile in general, rather than the history of solid accretion of the planet. 
After the disk gas completely dissipates, the planet is expected to cool and contract lowering the saturation limit.
Therefore, our reported metallicity may be considered an upper limit on the silicate abundances in planets close to their host star although it may be complicated by the thermal evolution of the envelope {\it during} pollution which we discuss in more detail in Section \ref{subsec:HighMetallicities}.

Saturation occurs for two reasons. First, the solid disk is densest close-in, resulting in high rates of accretion. Second, the saturation limit of SiO$_2$ at 0.1 au is low. This is the combined effect of high gas pressure within the radiative zone and low vapor pressure (of SiO$_2$) at envelope temperatures. The result is that the amount of material deposited in the atmosphere far exceeds the amount than can be held. The atmospheric abundance of SiO$_2$ therefore depends solely on the temperature-pressure profile of the envelope.

Gas pressure within the radiative zone is significantly lower in atmospheres at 3.5 au. Furthermore, the vapor pressure of H$_2$O is generally higher than the vapor pressure of SiO$_2$ (even in cooler atmospheres). Consequently, H$_2$O can be present in much higher mixing ratios with the atmospheric gas, up to an order of unity in mass, which can be reached at solid accretion rate $\dot{M} = 5\times 10^{-9}M_\oplus\,{\rm yr}^{-1}$.
As illustrated in Fig.~\ref{fig:In Situ Profiles}, H$_2$O mass fraction is higher in more gas-depleted disks (right column).
Gas density in the outer envelope is intrinsically related to the gas density of the disk in which the planet forms. The result is that for comparable levels of evaporation, H$_2$O mass fractions will be higher in more highly depleted disks simply by virtue of having lower H/He content. 

Impactor size also has a significant effect on the amount of pollution. Small impactors (solid lines in Fig.~\ref{fig:In Situ Profiles}) tend to enrich the atmosphere more efficiently than large impactors (dashed lines) by leading to approximately 2 orders of magnitude higher H$_2$O mass fraction. This is somewhat counterintuitive, since the mass vaporization rate (Eq.~\ref{eq:langmuir}) scales with surface area, and the amount of mass deposited by a single impactor is therefore roughly proportional to its radius squared. However, for a fixed total mass accretion rate, massive impactors are accreted less frequently. The number flux of accreted impactors scales inversely with their mass and subsequently their volume. The post-accretion metallicity is therefore inversely proportional to the radius of the accreted material. 

\subsection{Migration}

\begin{figure}
  \centering
\includegraphics[width=\columnwidth]{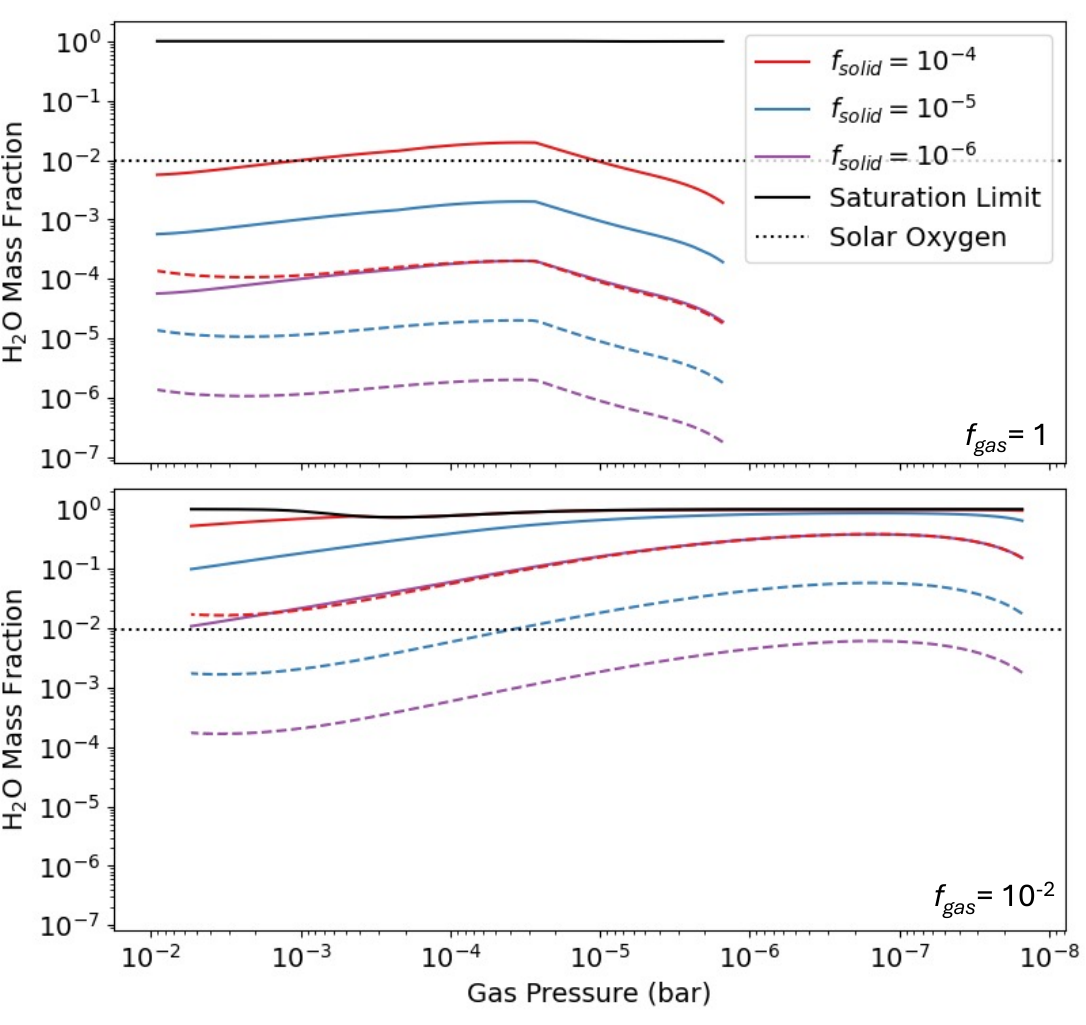}
\caption{Mass fractions of accreted H$_2$O relative to the to the total gas for planets migrating inwards from 3.5 au in disks of varying gaseous surface density ($f_{\rm gas} = 1$ at the top, $f_{\rm gas} = 10^{-2}$ at the bottom). Rapid migration in gas-heavy disks leads to significantly lower rate of pollution. All profiles are for dusty opacities. We indicate in horizontal line the oxygen mass fraction consistent with $Z=0.02$ and \citet{GN93}.}
\label{fig:Migration Profiles}
\end{figure}

We now consider pollution by migration. 
Ultimately, the rate of migration determines the timescale over which planets can accrete icy impactors, subsequently controlling the atmospheric H$_2$O levels.
A planet that migrates from 3.5 au in a non-depleted disk ($f_{\text{gas}} = 1$) passes the iceline (1 au) after $\sim$1.2$\times 10^4$ years, and therefore spends only a small fraction of the total 2 Myr accreting ice. The resultant low H$_2$O abundances are shown in the upper panel of Fig.~\ref{fig:Migration Profiles}. After the initial H$_2$O pollution, the planet spends the remainder of the disk lifetime in the inner disk accreting SiO$_2$, which is functionally the same as a planet accreting in situ at 0.1 au. While the exact SiO$_2$ profile will depend on the planet's thermal evolution, we expect silicate abundances similar to those in the upper panel of Fig.~\ref{fig:In Situ Profiles}. 

\begin{figure*}[]
  \centering
\includegraphics[width=1\textwidth]{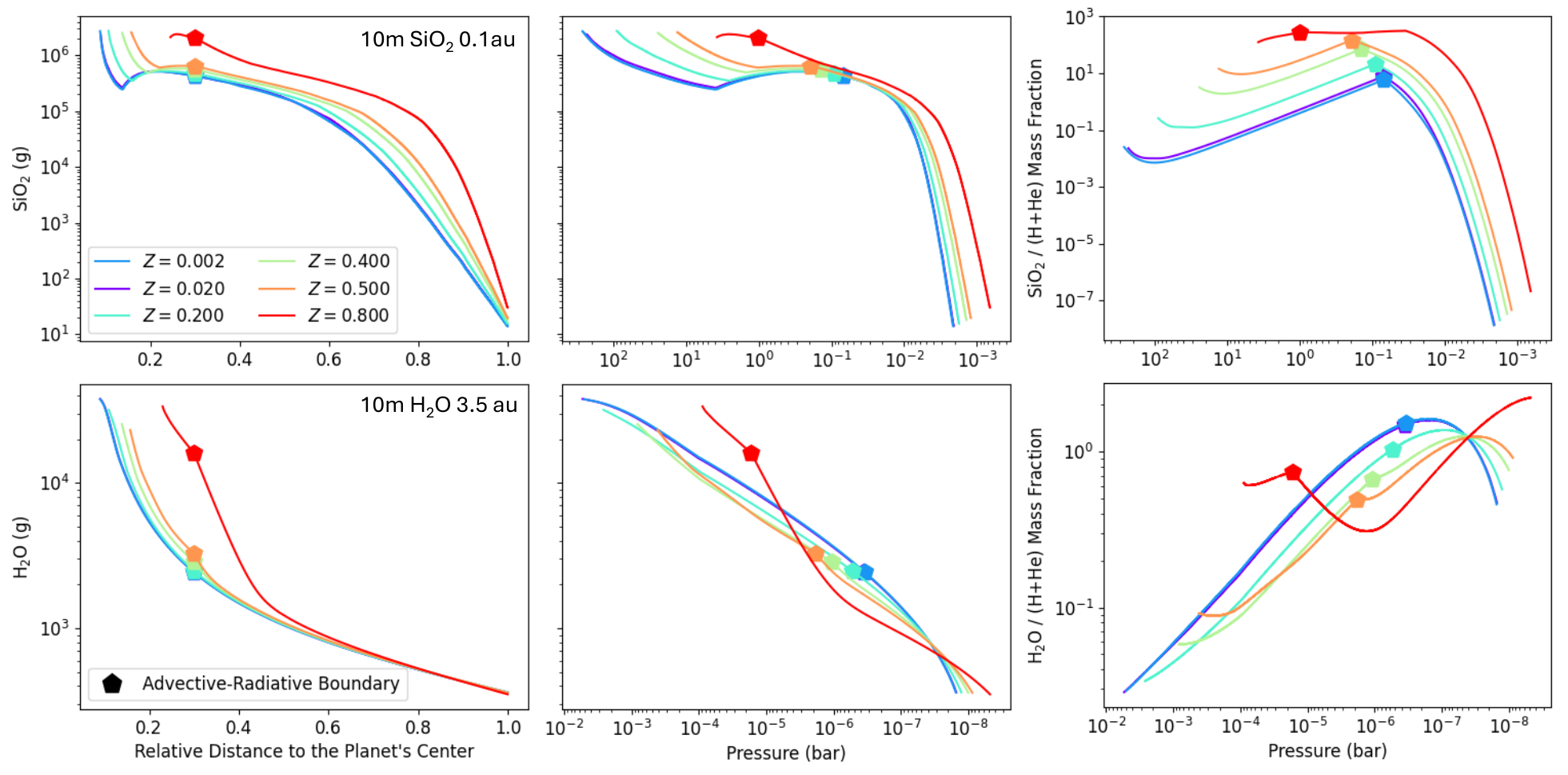}
\caption{We compare the radial profile of deposited mass of impactors (left and middle columns) and the final metallicity (deposited mass / gas mass only in H and He; right column) in atmospheres of varying initial metallicity $Z$. The left and middle panels show the mass deposited by a 10 m impactor as a function of the planet's radius and atmospheric pressure respectively. The upper panels depict SiO$_2$ at 0.1 au and the lower panels depict H$_2$O at 3.5 au. All atmospheric profiles have $f_{\rm gas}=10^{-2}$ and  $f_{\rm solid}=10^{-5}$. Pentacles indicate the advective-radiative boundary.}
\label{fig:Varying Metallicity}
\end{figure*}

A planet migrating in a disk depleted by $f_{\text{gas}}=10^{-2}$ takes 100 times longer to reach the iceline (i.e., comparable to the accretion time 2 Myr) and therefore spends more time accreting ice. Furthermore, the accretion occurs at intermediate distances where the solid surface density is higher at a given $f_{\rm solid}$, increasing the H$_2$O abundances even beyond those of in situ accretion at 3.5 au (for $f_{\text{gas}}=10^{-2}$; see bottom left panel of Fig.~\ref{fig:In Situ Profiles}) and approaching saturation limit at high $f_{\rm solid} = 10^{-4}$. 
These high H$_2$O abundances are shown in the lower panel of Fig.~\ref{fig:Migration Profiles}. The planet then stays in the inner disk accreting SiO$_2$ for the remainder of the disk lifetime ($\sim8 \times 10^5$ years), which is greater than the typical saturation timescale of a planet forming in situ at 0.1 au and so we expect the final SiO$_2$ abundance to be set by saturation limit that corresponds to the final temperature and pressure profile of the migrating planet, following its thermal evolution.

A planet in a highly depleted disk ($f_{\text{gas}}=10^{-4}$) does not migrate in a timescale of 2 Myr, so it is equivalent to that of in situ accretion at 3.5 au for $f_{\text{gas}}=10^{-4}$ (see lower left right panel of Fig.~\ref{fig:In Situ Profiles}). Because the planet does not pass the iceline, we do not consider the accretion of SiO$_2$. 

It is important to note that we do not account for the thermal feedback between the surrounding nebula and the protoplanetary atmosphere during migration. Planetary envelopes are expected to equilibrate to the local disk temperature and pressure over the disk lifetime, set to 2 Myr in our calculation.
In the case of $f_{\text{gas}}=1$, the timescale over which the planet accretes water is short enough for its thermal evolution to be negligible. Similarly, in the case of $f_{\text{gas}}=10^{-4}$, the total distance the planet migrates is short enough for its thermal evolution to be negligible. However, in the intermediate case ($f_{\text{gas}}=10^{-2}$) where the migration timescale is comparable to the disk lifetime, thermal evolution is likely non-negligible and the atmospheric H$_2$O abundances, as well as the saturation curves, may in fact be higher due to increased envelope temperatures as the planetary atmosphere equilibrates to hotter and denser gas. 

All H$_2$O abundance profiles for migrating planets are shown in Fig.~\ref{fig:Migration Profiles}. We note that variations between profiles stem from not only the variation in the amount of H$_2$O accreted but also the variation in atmospheric gas density. Just like our result for in-situ accretion, H$_2$O mass fractions are higher in more highly depleted disks under migration (see the bottom panel compared to the top panel) because of lower gas content in the upper layers of the atmosphere.

\section{Discussion}
\label{sec:discussion}

\subsection{Enrichment at High Metallicities}\label{subsec:HighMetallicities}
So far, we have not accounted for the changes in the temperature and pressure profile of the atmosphere as it becomes polluted. While simultaneously evolving the thermodynamical and compositional variations in time is beyond the scope of the paper, we evaluate the validity of our approach by comparing the amount of material that would be deposited by an impactor into atmospheres of varying metallicities. 

Higher atmospheric metallicty leads to higher opacity and therefore hotter atmosphere, resulting in higher amounts of evaporation at a fixed radial location from the planetary center (see left column of Fig.~\ref{fig:Varying Metallicity}). The increase in evaporated mass with higher metallicity is more pronounced throughout the entire envelope for SiO$_2$ because at 0.1 au, the upper envelope follows a non-isothermal profile due to higher opacity there as compared to 3.5 au (we adopt dust opacity $\kappa \propto T^2$ below log $(T/{\rm K}) < 2.7$) with all else equal, leading to temperatures that are globally higher throughout the entire envelope at higher $Z$ at 0.1 au. By contrast, at 3.5 au, lower opacities allow the outer envelope profiles to be isothermal with respect to the disk down to $\sim$0.1--0.5 $R_{\rm out}$ from $Z = 0.02$ to $Z=0.8$ below which the temperature quickly diverges, and so we observe higher H$_2$O deposition mass at higher $Z$ only below $\sim$0.5$R_{\rm out}$ at 3.5 au. The fact that the radial location of steep increase in envelope temperature is higher up in altitude at higher $Z$ gives rise to qualitatively different deposition mass profiles at $Z=0.8$ compared to lower $Z$'s in Figure \ref{fig:Varying Metallicity}. More precisely, at $Z=0.8$, unlike lower $Z$, the steep temperature increase already occurs in the outer advective layer (as reflected in the steep increase in deposited H$_2$O mass above the advective-radiative boundary at $Z=0.8$ in the lower left panel) such that the radiative zone is not isothermal.

At fixed pressure (which is the quantity that is more easily probed by observations), the evaporated mass profiles shift (see the middle column of Fig.~\ref{fig:Varying Metallicity}). The outermost temperature and density of our envelopes are anchored to that of the surrounding disk and the outermost pressure is then derived using the ideal gas EOS constructed in \citet{Lee_2014}. By fixing the nebular density to the same value for all $Z$, we obtain lower outermost pressure (by factors of order unity) for higher $Z$ because of higher mean molecular weight. At lower $Z$, the outer envelope remains more isothermal and therefore the pressure/density profile rises more in an exponential fashion whereas at higher $Z$, the outer envelope more closely follows a more adiabatic profile. With these differing profile shapes, the atmospheric pressure converges at $\sim$0.8$R_{\rm out}$ ($\sim$10$^{-2}$ bar at 0.1 au; $\sim$5$\times 10^{-7}$ bar at 3.5 au) below which they diverge such that the pressure is higher at higher $Z$ at a given radius. Such behavior in pressure translates to more divergence in evaporated mass at the outermost advective layers. Convergence is observed towards but above the radiative zone (i.e., the advective-radiative boundary is at 0.3$R_{\rm out}$ whereas the convergence is at 0.8$R_{\rm out}$) within which the evaporated mass diverges again as both temperature and pressure rise with $Z$. For H$_2$O, because its evaporated mass is largely equal across all $Z$ in the large radial extent of the advective layer, at a fixed pressure, the evaporated mass becomes lower at higher $Z$.

In terms of mass fraction (right column of Fig.~\ref{fig:Varying Metallicity}), we see that the amount of SiO$_2$ deposition (controlled by saturation) over H+He mass within our fiducial $Z=0.02$ envelope reaches a maximum value of $\sim$0.8 at $\sim$10$^{-1}$ bar (the advective-radiative boundary) and drops towards deeper layers. Such a configuration (heavier fluid on top of lighter fluid) is likely dynamically unstable promoting convection, potentially causing efficient mixing below $\sim$10$^{-1}$ bar, flattening the mass fraction profile there to the ratio of the total deposition mass of SiO$_2$ to total H+He mass in the radiative layer which is $\sim$0.2 set by the saturation limit. In terms of envelope profiles, the pressure, temperature, and density all vary within factors of $\sim$2--3 between $Z=0.02$ and $Z=0.2$; however, this small shift in thermodynamic properties lead to $\sim$an order of magnitude difference in the SiO$_2$/(H + He) mass fraction leading to an average value $\sim$0.95 in $Z=0.2$ atmosphere. Given the sensitivity of the saturation limit of quartz, we conclude that we cannot rule out the thermal effect of the deposition of SiO$_2$ and that a more careful calculation would likely lead to higher SiO$_2$ mass fraction than what we report in Section \ref{sec:results}. Even for 1 km impactors at $f_{\rm solid}=10^{-6}$ that do not reach the saturation limit, the average deposition mass fraction is $\sim$0.09 which would lead to a hotter envelope and higher rate of ablation than what is computed for a static envelope.

We now consider water deposition at 3.5 au. The water mass fraction can reach values of order unity at each layer for small impactors and high $f_{\rm solid}$. We see from Fig.~\ref{fig:Varying Metallicity}, however, that most of the mass fraction profiles lead to envelopes that are more enriched in water at higher altitudes which again would likely lead to efficient mixing, homogenizing the mass fraction profile to the ratio of the total mass of accreted solid to the total H+He mass in the radiative layer, $\sim$0.01--1 for 10m impactors and $\sim$0.0001--0.01 for 1km impactors from $f_{\rm solid}=10^{-6}$ to $10^{-4}$ (while we only show $f_{\rm solid}=10^{-5}$ in Figure \ref{fig:Varying Metallicity}, we have run our calculations at other $f_{\rm solid}$ whose results we report). For 1 km impactors, the extra average metallicity from pollution is less than the initial $Z/(1-Z) \sim 0.02$ so its impact on the thermal evolution of the envelope profile is likely negligible. For 10 m impactor, $f_{\rm solid} > 10^{-6}$, we cannot ignore thermal feedback from water deposition given that at 3.5 au, the total deposited mass becomes comparable to the H+He mass. As the closest analogue, we see that at $Z=0.8$, the bulk mass fraction of deposited H$_2$O over H+He gas would approach $\sim$0.6--6 ($f_{\rm solid}=10^{-5}$--$10^{-4}$) making a waterworld. Like the case at 0.1 au, we cannot rule out the thermal effect of the deposition of H$_2$O for small impactors at high $f_{\rm solid}$ and in those cases, our quoted atmospheric metallicity should be taken as a lower limit. We note however that post-disk cooling of the envelope will likely lower the metallicity back down. We defer a more careful calculation of concomitant thermal and deposition evolution to future work.

\begin{figure*}[]
  \centering
\includegraphics[width=1\textwidth]{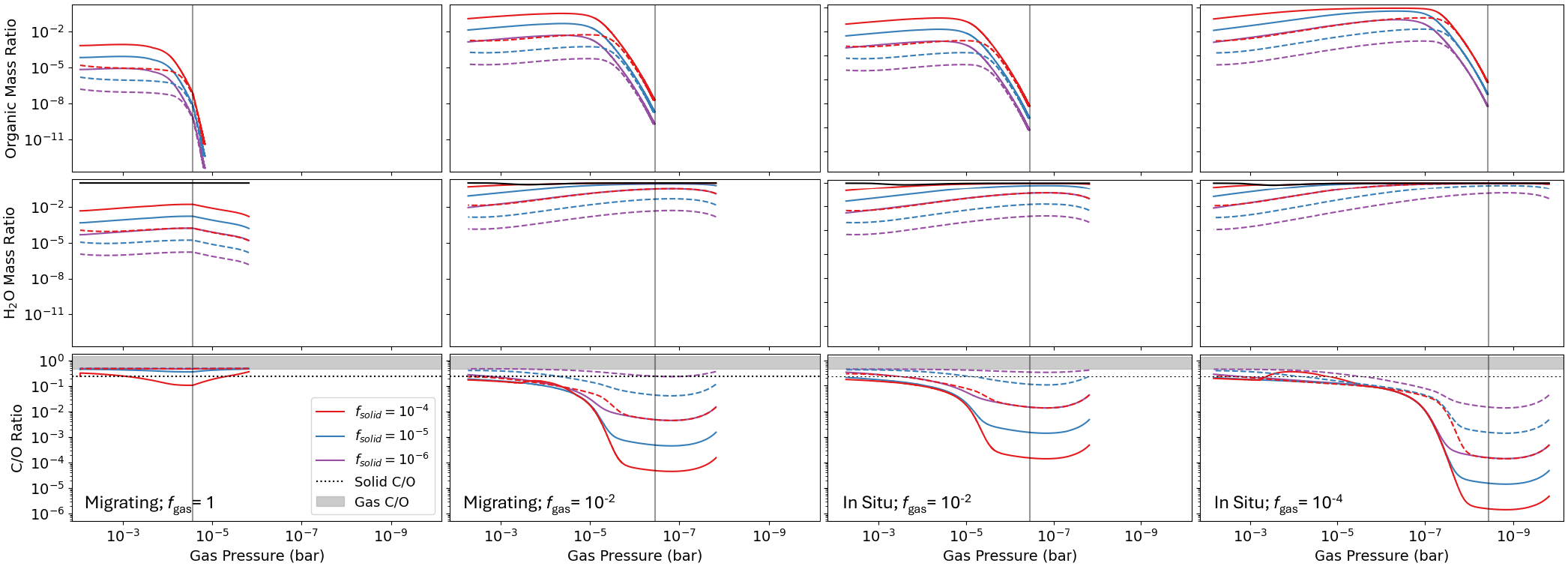}
\caption{Atmospheric mass fraction of refractory carbons (top) and water (middle) with the corresponding atmospheric carbon to oxygen mole ratio (bottom). Mass fractions are calculated for a solid disk consisting of 17.7\% refractory carbon and 82.3\% water ice. We assume atmospheric gas is initially of solar composition (C/O $\sim$ 0.48; \citealt{GN93}) and track how oxygen and carbon accreted in the form of H$_2$O and refractory carbons alter the C/O ratio. In the left-most panels, we show the C/O ratios of a planet migrating inwards from 3.5 au in a disk with $f_{\rm gas} = 1$ and $f_{\rm gas} = 10^{-2}$. In the right-most panels, we show the C/O ratios of a planet accreting in situ at 3.5 au in a disk $f_{\rm gas} = 10^{-2}$ and $f_{\rm gas} = 10^{-4}$. The vertical lines indicate the advective-radiative boundary, and all profiles terminate at the radiative-convective boundary. Maximum level of C/O ratio under pollution (i.e., the solid state C/O in the disk $\sim$0.24 representing the case where carbon refractory elements ablate just as easily as H$_2$O) is indicated in dashed line. We draw the disk gas C/O ratio with a horizontal bar from 1$\times$ to 3$\times$ the solar value with the latter representing the expected disk gas C/O ratio beyond the water but inside the CO$_2$ iceline (see text for detail). The bump in C/O ratio for $f_{\rm solid} = 10^{-4}$ in the second and fourth panels arises as the atmosphere becomes locally saturated with water.}
\label{fig:C to O ratios}
\end{figure*}

\subsection{Carbon to Oxygen Ratio}
We posit that the late-stage solid accretion of H$_2$O will lower the carbon to oxygen ratio of the atmospheres of envelopes beyond the iceline.

The exact C/O due to deposition depends on the sublimation of carbon carrying solids. Disk solids beyond the `soot' line (where all refractory carbon volatilizes) contain carbon in its refractory organic phase, which could sublimate off incoming impactors and deposit carbon into the atmosphere. Unlike SiO$_2$ and H$_2$O, refractory carbon sublimates into volatiles that do not recondense at the relevant temperatures. We therefore follow \citet{Chyba1990} and \citet{Li_2021} to calculate the rate of sublimation of refractory carbon using the Arrhenius equation
\begin{equation}
k(T) = A \exp{\left(-\frac{E_{\text{a}}}{RT}\right)}, 
\label{eq:refrac-subl}
\end{equation}
in lieu of Eq.~\ref{eq:langmuir}, where $A$ is in unit of frequency, $E_{\rm a}$ is the activation energy, and $R$ is the universal gas constant.

Organic matter within the interstellar medium is present as an insoluble, macromolecular material with a structure and composition similar to that of terrestrial kerogens \citep{Glavin_2018}, a common analogue to such refractory organics \citep{Chyba1990, Li_2021}. We adopt the kinetic parameters of type I kerogen measured by pyrolysis experiments of \citet{Burnham_1987} $E_{\text{a}} = 2.32\times 10 ^5$ J/mol and $A = 1.7\times 10^{14} \text{s}^{-1}$.\footnote{We choose type I kerogen because later in this section we introduce thermal conductivities for which we adopt the measured values of oil shales in the Green River formation whose properties are closer to that of type I rather than type II kerogen. Re-running our calculation using type II kerogen parameters lead to their ablation occurring at slightly lower temperature but ultimately the same organic mass ratio in the deeper atmosphere.}
Since the kinetic parameters are determined assuming a first order reaction, we approximate the sublimation rate as
\begin{equation} \label{eq:RefCarbAblation}
\dot{M}_s = - k(T_{\text{s}}) \cdot M_{\text{iso}},
\end{equation}
where $M_{\text{iso}} < M_s$ is the mass of the impactor that is heated to its surface temperature, $T_{\text{s}}$. 

To determine the depth to which the impactor of mass $M_{\rm s}$ equilibrates with $T_{\rm s}$ \citep{Brouwers_2018,Love_1991}, we approximate the steepness of the internal thermal gradient using the Biot number Bi defined as the ratio between the resistance to conduction and the resistance to convection at its surface,
\begin{equation}
\text{Bi} =  \frac{hL}{k_{\text{i}}},
\end{equation}
where $h=\sigma_{\text{sb}}|T_{\text{g}}  - T_{\text{s}}|^3$   is the characteristic heat transfer coefficient, $L = r/3$ is the characteristic lengthscale (for spherical impactors), and $k_{\text{i}}$ is the impactor’s thermal conductivity. 
Systems with a Biot number less than 0.1 can be treated as approximately isothermal. The depth and mass of the isothermal outer layer can therefore be approximated as
\begin{equation}
d_{\rm iso} = \text{min}\left(\frac{0.3 k_{\text{i}}}{\sigma_{sb} |T_{\text{g}}  - T_{\text{s}}|^3}, r_i\right)
\end{equation} and
\begin{equation}
M_{\text{iso}} = \frac{4}{3}\pi \rho_{\text{i}} \left(r_{\text{i}}^3 – (r_{\text{i}} – d_{\text{iso}})^3\right)
\end{equation}
respectively, where $\rho_{\rm i}=1.07\,{\rm g\,cm^{-3}}$ is the material density of type I kerogen \citep{Smith61}.

We evaluate $k_{\text{i}}$ according to 
\begin{equation}
k_{\text{i}}(T_{\text{s}}) = 1.311 - 6.29 \times 10^{-4} \cdot T_{\text{s}}, 
\end{equation}
based on measurements of kerogen-rich oil shales from \citet{Nottenburg_1978}. The expression above is valid for temperatures ranging from 298K--653K over which thermal conductivities range from 0.90 W/mK to 1.12 W/mK. We note that the thermal conductivities of oil shales reported in the literature range from 0.25 W/mK to 3 W/mK \citep[e.g.,][]{Gabova_2020}. Despite this broad uncertainty, we find that varying the thermal conductivity of the refractory carbon has little effect on the final C/O ratio, which differs by a maximum of $8\%$ between $k_{\text{i}} = 0.25$ W/mK and $k_{\text{i}} = 3$ W/mK.

The ablation of carbonaceous impactors is then evaluated following the same procedure as for SiO$_2$ and H$_2$O impactors but replacing Eq.~\ref{eq:langmuir} with Eq.~\ref{eq:RefCarbAblation} to obtain the amount of mass deposited into the atmosphere by refractory carbons. Although half of all carbon resides in the refractory phase in the interstellar medium \citep{Oberg2021}, the amount of carbon present in refractory phase in disk solids is typically lower due to thermal processing of disk solids. We estimate that only $\lesssim 25\%$ of carbon is in refractory carbon phase \citep{Chachan_2023}, based on measurements of least thermally processed carbon-rich meteorites in the solar system \citep{Bergin2015}. Further assuming that $50\%$ of disk oxygen is contained in water ice and another 25\% is tied to refractory elements such as silicon, this implies that the C/O of the disk solids is $\sim 1/3$ the stellar C/O and the C/O of the disk gas $\sim$3 the stellar value inside the CO$_2$ iceline. However, refractory oxides will not sublimate enough to contribute to the atmospheric oxygen abundance beyond the water snowline where the envelopes are cold.
Excluding their contribution, the C/O present in the solid disk is $\sim 1/2$ the stellar value ($\sim$0.24 for our solar composition). Assuming the refractory organic has a composition of $\text{C}_{100}\text{H}_{77}\text{O}_{14}\text{N}_{3.5}\text{S}_2$ based on the average composition of insoluble organic materials (IOM) from \cite{Bergin_2023},  we model the solid disk as being $82.3\%$ water and $17.7\%$ refractory organic by mass and scale the accretion rate of each substance accordingly. The resulting refractory carbon deposition fractions are shown in the upper panels of Fig.~\ref{fig:C to O ratios}. We find negligible sublimation of the refractory carbons in the outer envelope; however, above pressures of $\sim 10^{-8}$ bar, carbonaceous impactors reach sufficiently high temperatures to sublimate. In general, towards deeper atmosphere, we see a sharp increase in the refractory organic mass ratio followed by a plateau. The temperatures at which this plateau occurs is $\sim$450--530 K, which is 1--2 orders of magnitude smaller than the temperature that corresponds to $E_{\rm a}$ for kerogen in Eq.~\ref{eq:refrac-subl}. Instead, we find that this plateau occurs when the evaporative cooling overtakes frictional heating beyond which point $T_{\rm s}$ and by extension $\dot{M}_{\rm s}$ stabilizes.
In the inner envelope, the deposition fractions of the refractory carbon are comparable to those of water.

We calculate the C/O assuming an initial atmosphere with solar composition (C/O $\sim$ 0.48, \citealt{GN93} to be consistent with our atmospheric models; see Section \ref{subsec:atmosphere}) and updating according to the extra oxygen and carbon delivered from our computed H$_2$O and refractory carbon deposition fractions. The result from this calculation are shown in the bottom panels of Fig.~\ref{fig:C to O ratios}. In the outer envelope, the C/O depends strictly on the H$_2$O deposition and as such can vary by several orders of magnitude depending on the impactor size and solid disk depletion. Under marginal migration ($f_{\rm gas} = 10^{-2}$) and in situ condition, the C/O undergoes a steep increase below the advective-radiative boundary, where the refractory carbon begins to ablate. The variability between curves reduces, since high levels of H$_2$O deposition generally correspond to high levels of refractory carbon deposition. While the C/O in the outer envelope can be as low as $10^{-6}$, C/O within the inner envelope typically ranges from $\sim$0.05 to solar value.

\begin{figure*}[]
  \centering
\includegraphics[width=1\textwidth]{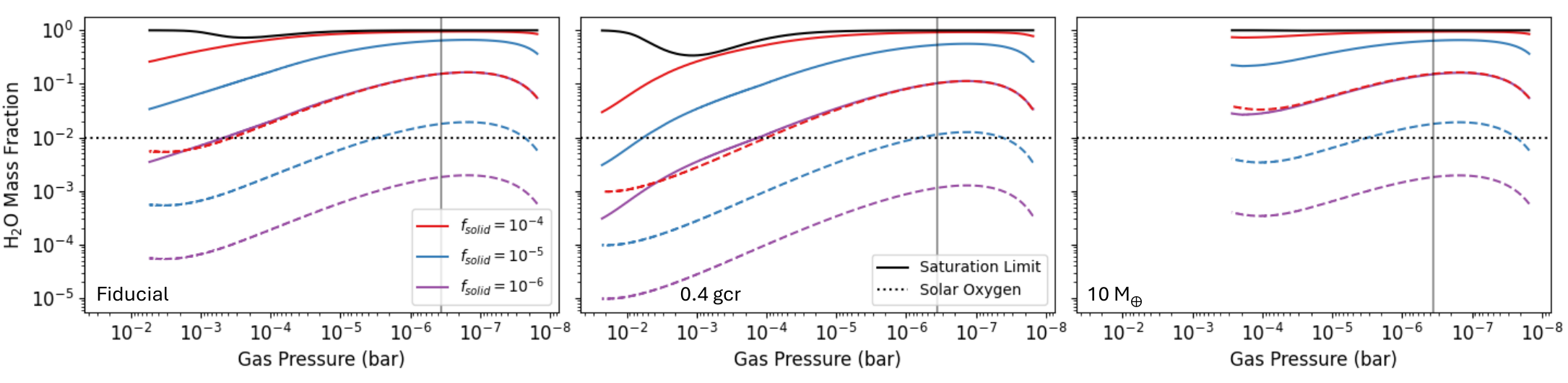}
\caption{Mass fraction of evaporated H$_2$O relative to the total gas. The left panel shows accretion into the fiducial model of a $5 M_{\oplus}$ core with a gas-to-core ratio (gcr) of $\sim$0.06. The center panel shows accretion into a planet with a gas-to-core ratio of $\sim 0.4$ (core mass fixed at $5 M_{\oplus}$) and the right panel shows accretion into a planet with a 10 $M_{\oplus}$ core (gas-to-core ratio fixed at 0.06). The vertical line marks the advective radiative boundary. All profiles are calculated for $f_{\rm gas} = 10^{-2}$ at an orbital distance of 3.5 au. The changes in deposition mass fraction with respect to gas-to-core mass ratio or core mass is negligible compared to the variations caused by different history of pollution.}
\label{fig:PlanetMass}
\end{figure*}

\subsection{Effect of Planet Mass}
\label{subsec:PlanetMass}

Among our solar system gas and ice giants, atmospheric metallicites appear to rise towards lower mass \citep{Atreya2020}. Such a trend is not easily observed in current exoplanet observations owing to large dispersion in measured metallicity for a given planet mass with a hint of such a dispersion increasing towards low mass planets \citep[e.g.,][their Figure 8]{Zhang2020}. Full scan of the parameter space from sub-Neptune to Jupiter-class objects is not the focus of our paper; instead, here, we deduce the projected direction of changes in the atmospheric metallicites with varying degrees of pollution around higher mass/radii sub-Neptune class objects by increasing the gas-to-core mass ratio to 0.3957 (at a fixed core mass; see middle panel of Fig.~\ref{fig:PlanetMass}) and increasing the core mass to 10 $M_\oplus$ (at a fixed gas-to-core mass ratio; see right panel of Fig.~\ref{fig:PlanetMass}).

Planets of higher gas-to-core mass ratio would appear significantly larger albeit at similar measured total mass \citep[e.g.,][]{Lopez2014}. In terms of ablation of impactors, we see that the mass fraction of deposited H$_2$O is generally lower compared to our fiducial case, albeit within factors of $\lesssim$10, with the most significant difference appearing in the deepest region of the radiative zone. The smaller mass fraction of H$_2$O in a more gas-heavy planet is largely because the total gas mass within each layer is higher and also because gas accretion is mediated by cooling \citep{Lee2015} and so higher gas-to-core mass ratio implies more cooled planet with lower entropy and lower temperature. The latter effect is minor however and we observe a similar degree of ablation in both cases. 

Finally, we consider H$_2$O accretion onto a planet with a $10 M_{\oplus}$ core with initial gas-to-core mass ratio $\sim$6\%. 
An envelope atop a more massive core is hotter with higher entropy leading to slightly more ablation of H$_2$O, but the difference is only within factors of $\sim$2. We observe that towards the radiative-convective boundary, the ablated H$_2$O mass fraction is lower in the fiducial case of 5$M_\oplus$ because this boundary appears deeper into the envelope compared to 10$M_\oplus$ and therefore the initial gas mass there at each layer is larger. The difference remains within an order of magnitude.

In summary, as long as we are looking at planets that would be classified as a sub-Neptune, we expect negligible change in atmospheric metallicity by pollution as a function of planet radii or mass as compared to the details of pollution (e.g., rate of pollution through $f_{\rm solid}$ or the rate of migration) that can cause orders of magnitude difference. In other words, scatter would overpower any discernible trend, unless we compare drastically different planet types (e.g., sub-Neptunes vs.~gas giants).

\section{Conclusion}
\label{sec:conclusions}

We have computed the metallicity profiles in the upper advective and radiative layers of a sub-Neptune class planet under pollution during the late stages of disk evolution where both solid and gas content are depleted (but not emptied). As infalling impactors fall through the planetary atmosphere, they heat up and ablate through friction and radiation. We summarize our results as follows:
\begin{enumerate}
    \item At 0.1 au, mass deposition by quartz impactors is always set by the saturation limit and therefore is controlled by the temperature and pressure of the atmosphere, with the exception of high $f_{\rm gas} \geq 10^{-2}$ and low $f_{\rm solid} \leq 10^{-6}$.
    \item Beyond the water ice line, mass deposition of water is set by the overall mass accretion rate and can vary by $\sim$4 orders of magnitude (2 orders of magnitude above the solar value) with lower level of mass deposition at lower rate of pollution (smaller $f_{\rm solid}$, larger impactor) and denser gas (higher $f_{\rm gas}$).
    \item Pollution by migration depends on the speed of migration. Rapid migration (early planet formation and high $f_{\rm gas}$) leads to low level of water mass fraction that would mostly lead to the solar value because not enough time is spent beyond the water ice line whereas marginal (i.e., migration time $\sim$ disk lifetime) and minimal migration would lead to high level of pollution as the planet spends more time beyond the ice line.
    \item Under marginal and minimal migration, concurrent accretion of pure water ice and refractory carbon solids can lead to C/O ratio that is as low as $\sim$10$^{-6}$ in the upper atmosphere but can increase up to $\gtrsim$0.1 in the deeper layers where refractory organics are expected to sublimate.
\end{enumerate}

In general, for sub-Neptunes, we expect post-formation pollution to erase all chemical memory of initial formation. In case of planets born and evolved inside the ice line, atmospheric silicate mass fraction is expected to simply reflect the thermal state of the atmosphere at the given time. In case of planets born outside the ice line, approximately 4 orders of magnitude variations in atmospheric metallicity (2 orders of magnitude above solar value) is expected depending on the rate of pollution that is controlled by the solid and gas mass content in the disk and the size of the impactors, more than enough to erase any trends that may have been emplaced during the initial formation of these planets. One interesting feature we observe in our calculations is that post-formation pollution can transform what was originally H/He-dominated sub-Neptune into a waterworld (defined as those with envelopes heavily enriched in water with atmospheric water mass fraction that approaches $\gtrsim$0.1--1) under marginal to minimal level of migration, consistent with the planet having formed relatively late when the disk gas is depleted by at least two orders of magnitude with respect to the solar value.

Our calculations are simplified to present one of the first steps towards assessing the effect of pollution on atmospheric composition of sub-Neptune class planets. Given that the expected metal-to-H/He mass fraction of SiO$_2$, H$_2$O, and refractory carbon can reach unity or higher, introducing more complexity that better reflects reality such as the thermal evolution of the atmosphere during and after pollution is warranted. Tracking post-pollution evolution of planetary atmosphere would also be necessary to determine the final observable atmospheric metallicity, which is a subject of future work.

\vspace{0.5cm}
We thank Antoine Bourdin and Mattias Lazda for providing some initial calculations of solid accretion rates. We also thank the anonymous referee for a report that helped improve the manuscript. We are grateful to Artem Aguichine, Ren\'{e} Doyon, Peter Gao, Chris Ormel, and Allona Vazan for helpful discussions. 
We recognize the support of the Digital Research Alliance of Canada (\url{https://alliancecan.ca/en}) and Compute Ontario (\url{https://www.computeontario.ca/}) through the use of their Graham cluster, which were crucial in the completion of this work. 
E.V. acknowledges the support of the Natural Sciences and Engineering Research Council of Cananada (NSERC) through the Undergraduate Student Research Awards (USRA).
Y.C. acknowledges support from the Natural Sciences and Engineering Research Council of Canada (NSERC) through the CITA National Fellowship and the Trottier Space Institute through the TSI Fellowship. 
V.S. acknowledges the support of the Natural Sciences and Engineering Research Council of Canada (NSERC) and of le Fonds de recherche du Québec – Nature et technologies (FRQNT). E.J.L. gratefully acknowledges support by NSERC, by FRQNT, by the Trottier Space Institute, and by the William Dawson Scholarship from McGill University.

\bibliography{pollution}{}
\bibliographystyle{aasjournal}

\end{document}